\documentclass[aps,prb,twocolumn,superscriptaddress,showpacs]{revtex4-1}
\usepackage{hyperref}

\usepackage[pdftex]{graphicx}
\usepackage{amssymb}
\usepackage{amsmath}
\usepackage[figure,table]{hypcap}
\usepackage{appendix}
\usepackage{bm}
\usepackage{bbm}

\newcommand{\ket}[1]{\left | #1 \right \rangle}
\newcommand{\bra}[1]{\left \langle #1\right |}
\newcommand{\hket}[1]{|| #1 \rangle}
\newcommand{\hbra}[1]{\langle #1||}
\newcommand{\tr}{\mbox{Tr}}

\begin{document}
\title{Dynamics of magnetic moments \\ coupled to electrons and lattice oscillations}
\date{\today}
\author{B. \surname{Mera}}
\email[E-mail: ]{bruno.mera@ist.utl.pt}
\author{V. R. \surname{Vieira}}
\affiliation{Centro de F\'{i}sica das Interac\c{c}\~{o}es Fundamentais, Instituto Superior T\'{e}cnico, Universidade T\'{e}cnica de Lisboa,\\
Av. Rovisco Pais 1, 1049-001 Lisboa, Portugal}
\author{V. K. \surname{Dugaev}}
\affiliation{Centro de F\'{i}sica das Interac\c{c}\~{o}es Fundamentais, Instituto Superior T\'{e}cnico, Universidade T\'{e}cnica de Lisboa,\\
Av. Rovisco Pais 1, 1049-001 Lisboa, Portugal}
\affiliation{Department of Physics, Rzesz\'{o}w University of Technology, al. Powsta\'{n}c\'{o}w Warszawy 6, 35-959 Rzesz\'{o}w, Poland}
\begin{abstract}
Inspired by the models of A. Rebei and G. J. Parker\cite{RebeiParker2003} and A. Rebei \textit{et. al.}\cite{RebeiParker2005}, we study a physical model which describes the behaviour of magnetic moments in a ferromagnet. The magnetic moments are associated to 3d electrons which interact with conduction band electrons and with phonons. We study each interaction separately and then collect the results assuming that the electron-phonon interaction can be neglected. For the case of the spin-phonon interaction, we study the derivation of the equations of motion for the classical spin vector and find that the correct behaviour, as given by the Brown equation for the spin vector and the Bloch equation, using the results obtained by D. A. Garanin\cite{Garanin1997} for the average over fluctuations of the spin vector, can be obtained in the high temperature limit. At finite temperatures we show that the Markovian approximation for the fluctuations is not correct for time scales below some thermal correlation time $\tau_{Th}$. For the case of electrons we workout a perturbative expansion of the Feynman-Vernon functional. We find the expression for the random field correlation function. The composite model (as well as the individual models) is shown to satisfy a fluctuation-dissipation theorem for all temperature regimes if the behaviour of the coupling constants of the phonon-spin interaction remains unchanged with the temperature. The equations of motion are derived.    
\end{abstract}
\pacs{-}
\maketitle
\section*{Introduction}
\label{sec:Introduction}
The discovery of the GMR effect in 1988, for which Peter Grunberg and Albert Fert were awarded the Nobel Prize in Physics in 2007, motivated scientific research of the magnetization dynamics at the scale of nanometers and led to the birth of a new field of research now called spintronics. Spintronic devices have nanometer-scale sizes, can operate in high frequencies ($\sim 1 \text{GHz}$) and have a wide range of applications which go from the creation of small dimension ($<1\mu\text{m}$) microwave frequency generators to the improvement of magnetic storage devices.\\
\indent To successfully design these devices one needs to develop the theoretical comprehension of magnetization dynamics at the appropriate scales. The complete understanding of magnetization dynamics at the nano-scale can only be achieved by theorizing from first principles and that implies a quantum mechanical treatment. In particular, if one wants to describe a spin system far from equilibrium, one needs to use the methods of quantum open systems far from equilibrium, namely the Keldysh\cite{Keldysh1964} or the Lindblad\cite{Lindblad1976} formalisms.\\
\indent It has been shown\cite{RebeiParker2003} that the linear coupling interaction of a spin with a bosonic bath allows for the existence of white noise in the equation of motion which, under some particular conditions regarding the density of states of the bath, adopts the form of the Landau-Lifshitz-Gilbert-Brown equation. Also, it has been shown that if the spin vector satisfies a Landau-Lifshitz equation supplemented with white noise, then the magnetic moment as the average over the fluctuations of the spin satisfies, in the limit of low temperatures, a Landau-Lifshitz equation and, in the limit of high temperatures, a Landau-Lifshitz-Bloch equation\cite{Garanin1997}. The collection of these results together with the known result of formal equivalence between Landau-Lifshitz and Landau-Lifshitz-Gilbert equations allows one to conclude that the interaction with phonons (or other bosonic bath satisfying certain conditions) can be responsible for the motion of magnetic moments as described by a Landau-Lifshitz-Bloch equation which, in fact, gives a good description of the physical situation at high temperatures\cite{Chubykalo2008}. A quantum field theoretical treatment of the s-d interaction of conduction electrons and spins\cite{RebeiParker2005} has shown that, in the semi-classical limit, the magnetization obeyed a generalized Landau-Lifshitz equation.\\
\indent The need of increasing the speed of storage of information in magnetic media and the limitations associated with the generation of magnetic field pulses by an electric current require the research for ways of controlling  the magnetization by other means than external magnetic fields. In 1996, subpicosecond demagnetization in ferromagnetic Nickel was achieved using a $60 \ fs$ laser in the experiments of Beaurepaire \textit{et. al.}\cite{Beaurepaire}. Manipulating magnetization with ultrashort (of the order of the femtosecond) laser pulses is now a major research challenge because at such time scales it might be possible to reverse the magnetization faster than within half a precessional period\cite{Kirilyuk2010}. Because of this, it is of fundamental importance to understand the time evolution of magnetic moments at high temperatures and time scales which are approaching the femtosecond.  Reviews on the state of art of ultrafast spin dynamics and prospects are given, for example, by  A. Kyrilyuk \textit{et. al.} \cite{Kirilyuk2010} and G. Zhang \textit{et. al.} \cite{Hillebrands}.\\
\indent The ultrashort laser pulses are expected to strongly couple lattice oscillations, that is, phonons, and/or conduction electrons to the spins. This suggests that we may consider an effective microscopic theory for the system in which the fundamental interactions are spin-phonon, spin-electron and electron-phonon type. In this work we introduce and study a theory to model this physical system in which, for the sake of simplicity, we will neglect the electron-phonon interaction.\\
\indent This paper is divided in three major sections. In the first section we consider a generalized version of the model of A. Rebei and G. J. Parker\cite{RebeiParker2003}, that modelled the interaction of a spin-$j$ and a bosonic bath, now written to describe the interaction between a spin-$j$ and a bath of phonons which are assumed to be spin-$1$. The second section considers a model of a ferromagnet, inspired by the work of A. Rebei \textit{et. al.}\cite{RebeiParker2005}, assuming the interaction of a magnetic system associated to 3d electrons, represented by a collection of spin-$j$ vectors, with conduction 4s electrons as the bath. In both sections we use a path integral representation for the coherent state matrix elements of the reduced density matrix associated with the system. In the case of the phonons, since we consider a linear interaction, they can be exactly integrated out. After obtaining the effective action appearing in the phase of the path integral we use a stationary phase approximation and obtain an equation of motion for the classical spin vector. We then introduce a random field in the equations of motion through a Hubbard-Stratonovich transformation in the path integral expression. The high temperature limit is discussed and the Landau-Lifshitz-Gilbert-Brown and Landau-Lifshitz-Bloch equations are recovered in this limit. The finite temperature case is considered. In the case of the conduction electrons, the same procedure of integrating out the bath is not possible due to the non linearity of the interaction. Nevertheless, we use a Hubbard-Stratonovich transformation and obtain an expansion for the effective action of the system which we truncate at the second order. The third section compiles the results of the first and the second sections in a single model assuming a quantum system composed of a spin-$j$ vector field interacting with a bath of spin-$\frac{1}{2}$ electrons and spin-$1$ phonons.\\
\indent Finally, we list our conclusions.
\section{spin-phonon theory}
\label{sec:I- spin-phonon theory}
In this model the system considered is, for simplicity, a single spin-$j$ which interacts, linearly, with a bath of spin-$1$ phonons which can have longitudinal and transverse polarizations.
\\ 
\indent The total Hamiltonian is written in the form
\begin{eqnarray}
\hat{\mathcal{H}}=\hat{\mathcal{H}}_{S}+\hat{\mathcal{H}}_{R}+\hat{\mathcal{H}}_{i}
\label{eq:QOS Hamiltonian}
\end{eqnarray}
We now introduce the quantum operators associated to the spin-$j$ particle $\hat{S}_{\alpha}$ which are Hermitian operators satisfying the angular-momentum commutation relations
\begin{eqnarray}
\left[\hat{S}_{\alpha},\hat{S}_{\beta}\right]=i\varepsilon_{\alpha\beta\gamma}\hat{S}_{\gamma},
\end{eqnarray}
where $\varepsilon_{\alpha\beta\gamma}$ is the Levi-Civita tensor density (repeated indexes are assumed to be summed from now on unless otherwise stated). We also introduce the creation and destruction operators associated with the phonon degrees of freedom $\hat{a}^{\dagger}_{\mu}(\textbf{k}), \ \hat{a}_{\mu}(\textbf{k})$ which are labelled by the polarization index $\mu=-1,\ 0$ or $\ 1$ and by a momentum index $\textbf{k}$. They satisfy the Weyl algebra
\begin{eqnarray}
\left[\hat{a}_{\mu}(\textbf{k}),\hat{a}^{\dagger}_{\nu}(\textbf{k}')\right]=\delta_{\mu\nu}\delta^{3}(\textbf{k}-\textbf{k}').
\end{eqnarray}
The spin operators and the phonon operators commute.\\
\indent The system Hamiltonian is considered to be
\begin{eqnarray}
\hat{\mathcal{H}}_{S}=-\hat{S}_{\alpha}H_{\alpha}=-\hat{\textbf{S}}\cdot\textbf{H}.
\end{eqnarray}
We write the phonon Hamiltonian in the form
\begin{eqnarray}
\hat{\mathcal{H}}_{R}=\sum_{\mu,\textbf{k}}\omega_{\mu}(\textbf{k})\hat{a}^{\dagger}_{\mu}(\textbf{k})\hat{a}_{\mu}(\textbf{k}),
\end{eqnarray}
where $\omega_{\mu}(\textbf{k})$ are the energy eigenvalues (transverse oscillations have the some frequency, that is, $\omega_{-1}(\textbf{k})=\omega_{1}(\textbf{k})$).\\
\indent We write the most general form for an Hermitian Hamiltonian coupling linearly the spin, in the macrospin approximation\cite{Xiao2005}, and the phonon operators,
\begin{eqnarray}
\hat{\mathcal{H}}_{i}=-\sum_{\textbf{k}}\left(H_{\alpha\mu}^{*}(\textbf{k})\hat{a}^{\dagger}_{\mu}(\textbf{k})\hat{S}_{\alpha}+H_{\alpha\mu}(\textbf{k})\hat{S}_{\alpha}\hat{a}_{\mu}(\textbf{k})\right).
\label{eq:Spin-phonon interaction}
\end{eqnarray}
The $SU(2)$ invariance is explicit if we ensure that the operator  
\begin{eqnarray}
\hat{H}_{\alpha}=\sum_{\textbf{k}}\left(H_{\alpha\mu}^{*}(\textbf{k})\hat{a}^{\dagger}_{\mu}(\textbf{k})+H_{\alpha\mu}(\textbf{k})\hat{a}_{\mu}(\textbf{k})\right)
\end{eqnarray}
transforms as a vector under a $SU(2)$-transformation associated with the spin-$j$ and consequently $\hat{\mathcal{H}}_{i}$ will behave as a scalar. We can then write the interaction Hamiltonian as
\begin{eqnarray}
\hat{\mathcal{H}}_{i}=-\hat{S}_{\alpha}\hat{H}_{\alpha}=-\hat{\textbf{S}}\cdot\hat{\textbf{H}}
\end{eqnarray}
\\
\indent Following Rebei and Parker\cite{RebeiParker2003}, we now consider the reduced density matrix associated with the system, that is,
\begin{eqnarray}
\hat{\rho}_{S}(t)=\tr _{R}\left\{\hat{\mathcal{U}}(t;t_{0})\mathcal{\rho}(t_{0})\hat{\mathcal{U}}^{\dagger}(t;t_{0})\right\},
\end{eqnarray}
where $\hat{\mathcal{U}}(t;t_{0})=T\exp\left(-i\int_{t_{0}}^{t}\hat{\mathcal{H}}(s)ds\right)$ in which $T$ is the time ordering operator. In many problems of magnetic systems one can consider that the bath relaxes much faster than the spin. Then the system and the bath, at time $t_{0}$, are decoupled. We make also the assumption that, at time $t_{0}$, the bath is in equilibrium with the spin. With these considerations, we have
\begin{eqnarray}
\hat{\rho}(t_{0})=\hat{\rho}_{S}(t_{0})\otimes\hat{\rho}_{R}(t_{0}), \\
\hat{\rho}_{R}(t_{0})=Z_{R}^{-1}\exp\left(-\beta\hat{\mathcal{H}}_{R}\right),\nonumber
\label{eq:Initial condition} 
\end{eqnarray}
where $Z_{R}=\tr\left\{\exp\left(-\beta\hat{\mathcal{H}}_{R}\right)\right\}$ and $\beta$ denote, respectively, the partition function and the inverse of the temperature of the bath at the initial instant of time.\\
\indent We will work the path integral expression for the reduced density matrix associated with the system using coherent state basis for the Hilbert spaces associated with our physical problem. For the phonons we use the so-called holomorphic representation in which the coherent states are defined by
\begin{eqnarray}
\hket{\alpha}=\prod _{\textbf{k}}{\exp\left(\hat{a}^{\dagger}_{\mu}(\textbf{k})\alpha_{\mu}(\textbf{k})\right)}\ket{0},
\end{eqnarray}
which satisfy the useful property
\begin{eqnarray}
\hat{a}_{\mu}(\textbf{k})\hket{\alpha}=\alpha_{\mu}(\textbf{k})\ket{0}.
\label{eq:Bosoncs}
\end{eqnarray}
These states provide a decomposition of the identity (of the reservoir) of the form,
\begin{eqnarray}
\hat{\mathbbm{1}}_{R}=\int d\mu(\alpha)\hket{\alpha}\hbra{\alpha},
\\ d\mu(\alpha)=\prod_{\textbf{k},\mu}\frac{d^{2}\alpha_{\mu}(\textbf{k})}{\pi}e^{-\alpha_{\mu}^{*}(\textbf{k})\alpha_{\mu}(\textbf{k})}.\nonumber
\end{eqnarray}
For the case of the spin-$j$ we use the so called spin coherent states \cite{Vieira1995-Ann, Vieira1995-Nucl, Perelomov},
\begin{eqnarray}
\ket{\textbf{S}}=(1+|\zeta(\textbf{S})|^{2})^{-j}\exp(\zeta(\textbf{S})\hat{S}_{-})\ket{jj},
\end{eqnarray} 
with
\begin{eqnarray}
\zeta(\textbf{S})=e^{i\varphi}\tan\frac{\theta}{2},\\ \textbf{S}=(S_{\alpha})=j(\sin\theta\cos\varphi,\sin\theta\sin\varphi,\cos\theta)^{T},
\end{eqnarray}
recall that $\hat{S}_{\pm}=\hat{S}_{1}\pm i\hat{S}_{2}$ and that $\ket{jj}$ is the highest weight vector of the representation of the group $SU(2)$ labelled by $j$. These states have the property,
\begin{equation}
\bra{\textbf{S}}\hat{S}_{\alpha}\ket{\textbf{S}}=S_{\alpha},
\label{eq:Spincs}
\end{equation}
and they provide a decomposition of the (system) identity operator in the form,
\begin{eqnarray}
\hat{\mathbbm{1}}_{S}=\int d\mu(\textbf{S})\ket{\textbf{S}}\bra{\textbf{S}},\\ d\mu(\textbf{S})=\frac{2j+1}{4\pi}\delta(\textbf{S}^2-j^2)d^{3}S.\nonumber
\end{eqnarray}
The properties \eqref{eq:Bosoncs} and \eqref{eq:Spincs} allow us to make a correspondence between classical and quantum quantities. We will consider path integral representations which make use of coherent state matrix elements which can be computed trivially, by the prescription $(\hat{a}_{\mu}^{\dagger}(\textbf{k}),\hat{a}_{\mu}(\textbf{k}),\hat{S}_{\alpha})\rightarrow(\alpha_{\mu}^{*}(\textbf{k}),\alpha_{\mu}(\textbf{k}),S_{\alpha})$. One can also define a holomorphic representation in the case of the spin coherent states and general rules for the insertion of operators can be derived\cite{Vieira1995-Ann}.
\\
\indent The spin coherent state matrix elements of the reduced density matrix at time $t$,
\begin{eqnarray}
&&\rho_{S}(\textbf{S}_{f},\textbf{S}_{i},t)= \bra{\textbf{S}_{f}}\hat{\rho}_{S}(t)\ket{\textbf{S}_{i}}\nonumber\\&=&\bra{\textbf{S}_{f}}\tr _{R}\left\{\hat{\mathcal{U}}(t;t_{0})\mathcal{\rho}(t_{0})\hat{\mathcal{U}}^{\dagger}(t;t_{0})\right\}\ket{\textbf{S}_{i}},
\end{eqnarray}
can be written in terms of those at time $t_{0}$ by using the propagator $J(\textbf{S}_{f},\textbf{S}_{i},t;\textbf{S}_{2},\textbf{S}_{1},t_{0})$, given by
\begin{eqnarray}
\rho_{S}(\textbf{S}_{f},\textbf{S}_{i},t)=&\int & d\mu(\textbf{S}_{1})d\mu(\textbf{S}_{2})J(\textbf{S}_{f},\textbf{S}_{i},t;\textbf{S}_{2},\textbf{S}_{1},t_{0})\nonumber\\ &\times & \rho_{S}(\textbf{S}_{1},\textbf{S}_{2},t_{0}).
\end{eqnarray}
The propagator has a path integral representation of the form
\begin{widetext}
\begin{eqnarray}
J(\textbf{S}_{f},\textbf{S}_{i},t;\textbf{S}_{2},\textbf{S}_{1},t_{0})=\int_{\textbf{S}_{1}}^{\textbf{S}_{f}} \mathcal{D}\mu(\textbf{S}_{1}')
\int_{\textbf{S}_{2}}^{\textbf{S}_{i}}\mathcal{D}\mu(\textbf{S}_{2}')\exp\left(i I[\textbf{S}_{1}',\textbf{S}_{2}']\right)\mathcal{F}[\textbf{S}_{1}',\textbf{S}_{2}']
\label{eq:Superpropagator}
\end{eqnarray}
\end{widetext}
where
\begin{eqnarray}
I[\textbf{S}_{1},\textbf{S}_{2}]&=& S_{\text{WZ}}[\textbf{S}_{1}]-S_{\text{WZ}}[\textbf{S}_{2}]+ \nonumber\\
&&+S_{S}[\textbf{S}_{1}]-S_{S}[\textbf{S}_{2}],
\label{eq:CTP WZ}
\end{eqnarray}
in which $S_{WZ}$ is the Wess-Zumino action,
\begin{eqnarray}
S_{\text{WZ}}[\textbf{S}] = j\int_{t_{0}}^{t} dt\int_{0}^{1}du \  \textbf{n}\cdot\left(\partial_{u}\textbf{n}\times\partial_{t}\textbf{n}\right),
\label{eq:WZ action}
\end{eqnarray}
where 
\begin{eqnarray}
\textbf{n}(u,t)=(\sin(u\theta)\cos\varphi ,\sin(u\theta)\sin\varphi ,\cos(u\theta) )^{T}
\label{eq:Homotopy map for the Wess-Zumino action}
\end{eqnarray}
is a map which continuously deforms the constant curve given by $\textbf{n}_{0}=(0,0,1)^{T}$ (the north pole) into the curve  $\textbf{n}(t)=\textbf{S}(t)/j$ through a great circle, a geodesic, in $S^{2}$. We see that the Wess-Zumino action, given by equation \eqref{eq:WZ action}, gives the area enclosed by the path traced by the spin vector and the two great circles from the north pole of the sphere to the endpoints of that path. We also have
\begin{eqnarray}
S_{S}[\textbf{S}] &=& -\int_{t_{0}}^{t}dt\bra{\textbf{S}(t)}\hat{\mathcal{H}_{S}}(t)\ket{\textbf{S}(t)}=\nonumber\\
&=& \int_{t_{0}}^{t}dt \ \textbf{S}(t)\cdot\textbf{H};
\end{eqnarray} 
and $\mathcal{F}[\textbf{S}_{1},\textbf{S}_{2}]$ is the Feynman-Vernon influence functional which can be expressed as
\begin{widetext}
\begin{eqnarray}
\mathcal{F}[\textbf{S}_{1},\textbf{S}_{2}]=\int d\mu(\alpha_{0})d\mu(\alpha_{1})d\mu(\alpha_{2}) \ k(\alpha_{0}^{\dagger},t;\alpha_{1},t_{0}|\textbf{S}_{1})\times Z_{R}^{-1}k(\alpha_{1}^{\dagger},-i\beta;\alpha_{2},0|0)\times k^{*}(\alpha_{0}^{\dagger},t;\alpha_{2},t_{0}|\textbf{S}_{2}),
\end{eqnarray}
\end{widetext}
where we have defined the kernel
\begin{eqnarray}
k(\alpha_{f}^{\dagger},t_{f};\alpha_{i},t_{i}|\textbf{S})&& \nonumber\\ &=&\int \mathcal{D}^{2}\alpha \exp\left[\frac{1}{2}\left(\alpha_{f}^{\dagger}\alpha(t_{f})+\alpha(t_{i})^{\dagger}\alpha_{i}\right)\right]  \nonumber\\
&&\times \exp\left[i\int_{t_{i}}^{t_{f}} dt \ L\right],
\label{eq:kernel k}
\end{eqnarray}
in which
\begin{eqnarray}
iL &=& \frac{1}{2}\left(\frac{d\alpha^{\dagger}}{dt}\alpha-\alpha^{\dagger}\frac{d\alpha}{dt}\right) -i[\mathcal{H}_{R}(\alpha^{\dagger}(t),\alpha(t)) \nonumber\\
&&+\mathcal{H}_{i}(\alpha^{\dagger}(t),\alpha(t),\textbf{S}(t))]
\end{eqnarray}
and
\begin{eqnarray}
\mathcal{H}_{R}= \frac{\hbra{\alpha(t)}\hat{\mathcal{H}}_{R}(t)\hket{\alpha(t)}}{\hbra{\alpha(t)}\hat{\mathbbm{1}}_{R}\hket{\alpha(t)}}, \\ 
\mathcal{H}_{i}=\frac{\bra{\textbf{S}(t)}\hbra{\alpha(t)}\hat{\mathcal{H}}_{i}(t)\hket{\alpha(t)}\ket{\textbf{S}(t)}}{\hbra{\alpha(t)}\hat{\mathbbm{1}}_{R}\hket{\alpha(t)}}. 
\end{eqnarray}
Note that, in the above formulas, we have adopted the vector notation $\alpha=(\alpha_{\mu}(\textbf{k}))$ and $\alpha^{\dagger}=(\alpha_{\mu}^{*}(\textbf{k}))$
The Feynman-Vernon functional can be computed exactly because the integrals are all Gaussian. After some algebra, we obtain
\begin{widetext}
\begin{eqnarray}
\mathcal{F}[\textbf{S}_{1},\textbf{S}_{2}] = \exp\left\{-i\left[\int_{t_{0}}^{t}\int_{t_{0}}^{t}dt_{1}dt_{2}\left(S_{1,\alpha}(t_{1})\ S_{2,\alpha}(t_{1})\right)\sigma_{3}\hat{G}_{\alpha\beta}(t_{1}-t_{2})\sigma_{3}\left(\begin{array}{c}
S_{1,\beta}(t_{2})\\
S_{2,\beta}(t_{2})
\end{array}\right)\right]\right\},
\end{eqnarray}
where,
\begin{eqnarray}
i\hat{G}_{\alpha\beta}(t) &=& \sum_{\textbf{k},\mu} H_{\alpha\mu}^{*}(\textbf{k})e^{-i\omega_{\mu}(\textbf{k})t}\nonumber\\ && \times \left(\begin{array}{cc}
\left(n(\omega_{\mu}(\textbf{k}))+1\right)\theta(t)+n(\omega_{\mu}(\textbf{k}))\theta(-t) & n(\omega_{\mu}(\textbf{k}))\\
n(\omega_{\mu}(\textbf{k}))+1 & \left(n(\omega_{\mu}(\textbf{k}))+1\right)\theta(-t)+n(\omega_{\mu}(\textbf{k}))\theta(t)
\end{array}
\right)H_{\beta\mu}(\textbf{k}),
\end{eqnarray}
\end{widetext}
in which $n(\omega)=(e^{\beta\omega}-1)^{-1}$ is the Bose-Einstein distribution function. If we introduce the fields associated with the Keldysh representation
\begin{eqnarray}
\left(\begin{array}{c}
S_{\alpha}(t)\\
D_{\alpha}(t)
\end{array}
\right)=\left(
\begin{array}{c}
\frac{1}{2}\left(S_{\alpha ,+}(t)+S_{\alpha ,-}(t)\right)\\
S_{\alpha ,+}(t)-S_{\alpha ,-}(t)
\end{array}
\right),
\label{eq:Keldysh rep}
\end{eqnarray}
we know that the field $\textbf{S}(t)$ is associated with the classical spin and $\textbf{D}(t)$ is associated with quantum and thermal fluctuations. The Feynman-Vernon functional in terms of these fields is
\begin{widetext}
\begin{eqnarray}
\mathcal{F}[\textbf{S},\textbf{D}] = \exp\left\{-i\int_{t_{0}}^{t}dt_{1}\int_{t_{0}}^{t}dt_{2}\left(\sqrt{2}S_{\alpha}(t_{1})\ \frac{1}{\sqrt{2}}D_{\alpha}(t_{1})\right)\tilde{G}^{\alpha\beta}(t_{1}-t_{2})\left(\begin{array}{c}
\sqrt{2}S_{\beta}(t_{2})\\
\frac{1}{\sqrt{2}}D_{\beta}(t_{2})
\end{array}
\right)\right\},
\end{eqnarray}
where,
\begin{eqnarray}
i\tilde{G}_{\alpha\beta}(t) = \sum_{\textbf{k},\mu}H_{\alpha\mu}^{*}(\textbf{k})e^{-i\omega_{\mu}(\textbf{k})t}\left(\begin{array}{cc}
0 & -\theta(-t)\\
\theta(t) & 1+2n(\omega_{\mu}(\textbf{k}))) 
\end{array}
\right)H_{\beta\mu}(\textbf{k}).
\end{eqnarray}
\end{widetext}
In the stationary phase approximation one considers the variation of the action in the phase of the path integral to be zero. The equations of motion which follow from taking this procedure are
\begin{eqnarray}
\dot{\textbf{D}}(t) =\textbf{S}(t)\times\frac{\delta S_{\text{eff}}}{\delta \textbf{S}(t)}+ \textbf{D}(t)\times\frac{\delta S_{\text{eff}}}{\delta \textbf{D}(t)},\\
\dot{\textbf{S}}(t) =\textbf{S}(t)\times\frac{\delta S_{\text{eff}}}{\delta \textbf{D}(t)}+\frac{1}{4} \textbf{D}(t)\times\frac{\delta S_{\text{eff}}}{\delta \textbf{S}(t)}.
\label{eq:Generalized LLG equation}
\end{eqnarray}
where
\begin{eqnarray}
S_{\text{eff}}[\textbf{S},\textbf{D}]=\int_{t_{0}}^{t}dt\ \textbf{D}(t)\cdot\textbf{H}-i\log\mathcal{F}[\textbf{S},\textbf{D}].
\end{eqnarray}
We notice that there is no quadratic term involving two $\textbf{S}(t)$ fields. By performing a Hubbard-Stratonovich transformation to the quadratic term in the action which couples two $\textbf{D}(t)$ fields, as in \cite{RebeiParker2003}, we obtain a random field in the equations of motion, $\bm{\xi}(t)$. The real part of the correlation function of this  random field is
\begin{eqnarray}
\text{Re}\left\{\langle\xi_{\alpha}(t_{1})\xi_{\beta}(t_{2})\rangle \right\}&=&\text{Re}\left\{\langle\xi_{\alpha}(0)\xi_{\beta}(t_{1}-t_{2})\rangle\right\} \nonumber\\
&=&\text{Re}\{\sum_{\textbf{k},\mu}H_{\alpha\mu}^{*}(\textbf{k})e^{-i\omega_{\mu}(\textbf{k})(t
_{1}-t_{2})}\nonumber\\
&&\times\left(1+2n(\omega_{\mu}(\textbf{k}))\right)H_{\beta\mu}(\textbf{k})\}.
\end{eqnarray} 
The last expression can be re-written as
\begin{eqnarray}
\text{Re}\{\sum_{\textbf{k},\mu}\int_{0}^{\infty}d\omega\delta(\omega-\omega_{\mu}(\textbf{k}))&& H_{\alpha\mu}^{*}(\textbf{k})H_{\beta\mu}(\textbf{k}) e^{-i\omega_{\mu}(\textbf{k})(t
_{1}-t_{2})} \nonumber\\ &&\times \coth\left(\frac{\beta\omega_{\mu}(\textbf{k})}{2}\right)\}
\end{eqnarray}
Let us assume that $H_{\alpha\mu}(\textbf{k})=H_{\alpha}(\omega_{\mu}(\textbf{k}))$. In this case, we can further write, interchanging the sum and the integral and using the property $f(\omega_{k})\delta(\omega_{k}-\omega)=f(\omega)\delta(\omega_{k}-\omega)$ of the delta distribution, we find
\begin{eqnarray}
\text{Re}\{\int_{0}^{\infty}d\omega\sum_{\textbf{k},\mu}\delta(\omega-\omega_{\mu}(\textbf{k}))  H_{\alpha}^{*}(\omega) && H_{\beta}(\omega) e^{-i\omega(t
_{1}-t_{2})}\nonumber\\ &&\times \coth\left(\frac{\beta\omega}{2}\right)\}
\end{eqnarray}
or, equivalently,
\begin{eqnarray}
\text{Re}\{\int_{0}^{\infty}d\omega \ \rho(\omega) H_{\alpha}^{*}(\omega) H_{\beta}(\omega) && e^{-i\omega(t
_{1}-t_{2})} \nonumber\\
&&\times \coth\left(\frac{\beta\omega}{2}\right)\}
\end{eqnarray}
where we have defined the density of states
\begin{eqnarray}
\rho(\omega)=\sum_{\textbf{k},\mu}\delta(\omega-\omega_{\mu}(\textbf{k})).
\end{eqnarray}
In the high temperature limit, the last expression becomes
\begin{eqnarray}
2k_{B}T \ \text{Re}\{\int_{0}^{\infty} \rho(\omega) \frac{H_{\alpha}^{*}(\omega)H_{\beta}(\omega)}{\omega}e^{-i\omega(t
_{1}-t_{2})}d\omega \}. 
\end{eqnarray}
If we consider a linear dispersion relation for the longitudinal and transverse phonons,
\begin{equation}
\omega_{0}(\textbf{k})=c_{l}k,\ \omega_{\pm 1}(\textbf{k}) =c_{t}k,
\end{equation}
and if we take the continuum limit for the bath, we arrive at the density of states
\begin{eqnarray}
\rho(\omega) &=& \sum_{\textbf{k},\mu}\delta(\omega-\omega_{\mu}(\textbf{k})) \nonumber\\
&=& V\sum_{\mu}\int\frac{d^{3}k}{(2\pi)^{3}}\delta(\omega-\omega_{\mu}(\textbf{k}))=\frac{V}{2\pi^{2}}\left(\frac{1}{c_{l}^{3}}+\frac{2}{c_{t}^{3}}\right)\omega^{2}\nonumber\\
&=& \rho_{0} \left(\frac{\omega}{\omega_{0}}\right)^{2},
\end{eqnarray}
where $V$ is the volume of the reservoir and $\rho_{0}=\rho (\omega_{0})$ is the density of states evaluated at some frequency $\omega_{0}$ taken as reference. This means that if we want to recover the Gaussian random field in the limit of high temperatures we must ensure that
\begin{eqnarray}
\text{Re}\{H_{\alpha}^{*}(\omega)H_{\beta}(\omega)\}\propto \frac{1}{\omega}.
\end{eqnarray}
In particular, one must have,
\begin{eqnarray}
\frac{\text{Re}\{ H_{\alpha}^{*}(\omega)H_{\beta}(\omega)\}\rho(\omega)\pi}{\omega}=\alpha_{G}\delta_{\alpha\beta},
\end{eqnarray}
so that,
\begin{eqnarray}
\text{Re}\left\{\langle\xi_{\alpha}(t)\xi_{\beta}(t')\rangle\right\}=\delta_{\alpha\beta}2\alpha_{G} k_{B}T\delta(t-t').
\end{eqnarray}
The constant $\alpha_{G}$ is a parameter which gives the intensity of the Gilbert damping term in the equation of motion.
The last condition constraints the form of $H_{\alpha}(\omega)$. If we define
\begin{eqnarray}
z_{\alpha}=\left(\frac{\pi\rho_{0}}{\alpha_{G}\omega_{0}^{2}}\right)^{1/2}\omega^{1/2}H_{\alpha}(\omega) .
\end{eqnarray}
then the above condition reads,
\begin{eqnarray}
z_{\alpha}^{*}z_{\beta}+z_{\beta}^{*}z_{\alpha}=2\delta_{\alpha\beta},
\end{eqnarray}
which means that
\begin{eqnarray}
z_{\alpha}=e^{i\varphi_{\alpha}}
\end{eqnarray}
and that the phases must satisfy
\begin{eqnarray}
\varphi_{\alpha}=\varphi_{\beta}+\left(2n+1\right)\frac{\pi}{2},
\end{eqnarray}
for all $\alpha$ different from $\beta$ and with $n$ being an integer. The form of $H_{\alpha}(\omega)$ is, thus, constrained to be
\begin{eqnarray}
H_{\alpha}(\omega)=\left(\frac{\pi\rho_{0}}{\alpha_{G}\omega_{0}^{2}}\right)^{-1/2}\omega^{-1/2}e^{i\varphi_{\alpha}},
\end{eqnarray}
if the familiar fluctuation-dissipation theorem is to be satisfied.\\
\indent Assuming the above dependence for $H_{\alpha}(\omega)$ for arbitrary temperatures, we find that the noise correlation function is written in the form
\begin{eqnarray}
\text{Re}\left\{\langle\xi_{\alpha}(t)\xi_{\beta}(0)\rangle\right\} && \nonumber\\
&=&\alpha_{G}\delta_{\alpha\beta}\int_{0}^{\infty}\frac{d\omega}{\pi}\ \omega \coth\left(\frac{\beta\omega}{2}\right)\cos\omega t \nonumber\\ &=& \alpha_{G}\delta_{\alpha\beta} \varphi(t),
\end{eqnarray}   
where we have defined the generalized function $\varphi$
\begin{eqnarray}
\varphi(t)= \int_{0}^{\infty}\frac{d\omega}{\pi} \ \omega \coth\left(\frac{\beta\omega}{2}\right)\cos\omega t,
\label{eq:Def of varphi}
\end{eqnarray}
which completely characterizes the noise correlation function. The study of this generalized function or distribution can be found, for instance, in the book of C. W. Gardiner and P. Zoller on \textit{Quantum Noise}\cite{Gardiner2}. The above integral can be expressed as
\begin{eqnarray}
\varphi(t) &=& \int_{0}^{\infty}\frac{d\omega}{\pi}\  \left[\coth\left(\frac{\beta\omega}{2}\right)\omega - \omega \right]\cos\omega t \nonumber\\ &&+\frac{1}{2}\left(\int_{-\infty}^{\infty}\frac{d\omega}{2\pi}\ |\omega |e^{-i\omega t}\right).
\end{eqnarray}
The first integral is convergent and can be computed. The result is
\begin{eqnarray}
&&\int_{0}^{\infty}\frac{d\omega}{\pi}\  \left[\coth\left(\frac{\beta\omega}{2}\right)\omega - \omega \right]\cos\omega t \nonumber\\
&=&\frac{1}{2\pi}\left\{\frac{1}{t^2}-\left(\frac{\pi}{\beta}\right)^{2}\text{cosech}^{2}\left[\left(\frac{\pi}{\beta}\right)t\right]\right\}.
\end{eqnarray}
The other integral can be computed if we regularize it with an exponential cut off,
\begin{eqnarray}
\int_{-\infty}^{\infty}\frac{d\omega}{2\pi}\ |\omega |e^{-i\omega t} &\rightarrow & \int_{-\infty}^{\infty}\frac{d\omega}{2\pi}\ |\omega |e^{-i\omega t}e^{-\lambda \omega} \nonumber\\
&=&\frac{1}{\pi}\frac{\lambda^{2}-t^{2}}{\left(\lambda^{2}+t^{2}\right)^{2}}, \ \lambda\rightarrow 0^{+}.
\end{eqnarray}
Formally we can, thus, write,
\begin{eqnarray}
\varphi(t)=-\frac{\pi}{2\beta^{2}}\text{cosech}^{2}\left[\left(\frac{\pi}{\beta}\right)t\right]
\end{eqnarray}
The correlation exhibits a thermal correlation time arising from the behaviour of $\varphi$ in the limit $|t|\rightarrow \infty$ which is
\begin{eqnarray}
\varphi(t) &\rightarrow & -2\pi\left(k_{B}T\right)^{2}\exp\left(-2\pi k_{B}T |t|\right)\nonumber\\
&=& -2\pi\left(k_{B}T\right)^{2}\exp\left(-\frac{|t|}{\tau_{\text{Th}}}\right)
\end{eqnarray}
where $\tau_{\text{Th}}=\left(2\pi k_{B}T\right)^{-1}$. In the case of arbitrarily small temperatures, this correlation time becomes very large (in fact, the correlation function becomes inverse quadratic in $|t|$). 
\\
\indent It is interesting to notice that the long time behaviour of the correlations is negative. One must study the full expression for $\varphi$ for finite cut off $\lambda$ to understand what is happening. We have
\begin{eqnarray}
\int_{-\infty}^{\infty} dt \ \text{Re}\left\{\langle\xi_{\alpha}(t)\xi_{\beta}(0)\rangle\right\} =2\alpha_{G}k_{B}T\delta_{\alpha\beta}
\label{eq:Fdt ph}
\end{eqnarray}
and what really happens is that, in fact, the cut-off dependent positive term is usually larger than the asymptotic negative term, except at zero temperature when the two effects cancel each other.\\
\indent This whole discussion can be simplified if one understands that the integral in the definition of $\varphi$, \eqref{eq:Def of varphi}, can be written as
\begin{eqnarray}
\varphi(t)=\alpha_{G}k_{B}T\frac{d}{dt}\coth\left[\pi k_{B}T t\right],
\end{eqnarray}
but then one must take into account, as pointed by G. W. Ford and R. F. O' Connel\cite{Ford-1993}, that the correct formula for the derivative of the hyperbolic cotangent is, in fact,
\begin{eqnarray}
\frac{d\coth x}{dx}=-\text{csch}^{2}x +2\delta(x).
\end{eqnarray}
The results are identical in both approaches.\\
\indent The existence of a thermal correlation time, which in standard units is $\tau_{\text{Th}}=h/k_{B}T= (1.27\times 10^{-12}\ s)/ T$, implies that the approximation of a Markovian description for the random field is only valid for times scales which are longer than this one or in the limit of large temperatures. In the case of the problem of controlling magnetization using ultrashort laser pulses which end up exciting lattice oscillations it might not be a good approximation to consider the process as Markovian. Equation \eqref{eq:Fdt ph} is remarkable because it is a manifestation of a fluctuation-dissipation theorem and it is, in fact, valid in all temperature regimes. \\
\indent We also note that, with the assumptions we've done regarding the constants $H_{\alpha}(\omega)$, we will always recover the Landau-Lifshitz-Gilbert form of the equations of motion in the limit $\textbf{D}\rightarrow 0$. To see this we just need to note that
\begin{eqnarray}
\frac{\delta S_{\text{eff}}}{\delta\textbf{D}(s)}\bigg|_{\textbf{D}(s)=0}=\int_{t_{0}}^{t}dt'\textbf{G}(s-t')\textbf{S}(t')
\end{eqnarray}
where the matrix elements of $\textbf{G}$ are
\begin{eqnarray}
G_{\alpha\beta}(t) &=& 2i\sum_{\textbf{k},\mu} \text{Re}\left\{H_{\alpha\mu}^{*}(\textbf{k})H_{\beta\mu}(\textbf{k})\right\} \nonumber \\ && \times e^{-i\omega_{\mu}(\textbf{k})t}\theta(-t).
\end{eqnarray}
If we use all the assumptions we made, the last expression becomes
\begin{eqnarray}
G_{\alpha\beta}(t)=2\alpha_{G}\delta_{\alpha\beta}\frac{d}{dt}\left[\delta(t)\right]\theta(-t),
\end{eqnarray}
which yields the result
\begin{eqnarray}
\frac{\delta S_{\text{eff}}}{\delta\textbf{D}(s)}\bigg|_{\textbf{D}(s)=0}=\alpha_{G}\dot{\textbf{S}}(s),
\end{eqnarray}
so that the equation of motion for the spin becomes always the Landau-Lifshitz-Gilbert equation in this limit. In the case of high temperatures we also recover the Landau-Lifshitz-Gilbert-Brown equation with the random field and due to the results of D. A. Garanin\cite{Garanin1997} we are able to recover the Landau-Lifshitz-Bloch equation for the average over fluctuations of the spin. To do that we neglect a term in the equation of motion which couples two spin fields and a random field. This is valid in the limit of small fluctuations. 
\section{Spin-electron theory}
\label{sec:II- Spin-electron theory}
\indent In this model the system can be viewed as a set of spins in a lattice, each spin written in the form of a general spin-$j$ representation of the $SU(2)$ group, modelling 4d-type electrons in a magnetic medium. We consider the reservoir to be composed of conduction electrons. The interaction Hamiltonian is an s-d  type interaction of the conduction electrons with the spin vector field.\\
The model Hamiltonian we consider here is again of the form of \eqref{eq:QOS Hamiltonian}, where
\begin{eqnarray}
\hat{\mathcal{H}}_{S}=-\sum_{i}\textbf{H}_{i}\cdot\hat{\textbf{S}}_{i}-\frac{1}{2}\sum_{ij}J_{ij}\hat{\textbf{S}}_{i}\cdot\hat{\textbf{S}}_{j}, \ J_{ij}\geq 0,
\end{eqnarray}
given in the momentum representation by
\begin{eqnarray}
\hat{\mathcal{H}}_{S} &=&-\sum_{\textbf{k}}\textbf{H}(-\textbf{k})\cdot\hat{\textbf{S}}(\textbf{k})\nonumber \\ &&-\frac{1}{2}\sum_{\textbf{k}}J(\textbf{k})\hat{\textbf{S}}(-\textbf{k})\cdot\hat{\textbf{S}}(\textbf{k}),\end{eqnarray}
The other terms are given by
\begin{eqnarray}
\hat{\mathcal{H}}_{R}&=&\sum_{\textbf{k},\alpha}\epsilon(\textbf{k})\hat{c}_{\alpha}^{\dagger}(\textbf{k})\hat{c}_{\alpha}(\textbf{k})\nonumber\\ &&-\lambda\sum_{\textbf{k}}\hat{\textbf{s}}(-\textbf{k})\cdot\hat{\textbf{S}}(\textbf{k})\nonumber\\
&&-\sum_{\textbf{k}}\hat{\textbf{s}}(-\textbf{k})\cdot\textbf{h}(\textbf{k}),
\end{eqnarray}
where $\epsilon(\textbf{k})$ are the energies of the conduction electrons, $\textbf{h}(\textbf{k})$ denotes the magnetic field felt by the conduction electrons and $\hat{\textbf{s}}(\textbf{k})$ denotes the composite operator
\begin{eqnarray}
\hat{\textbf{s}}(\textbf{k})=\frac{1}{2}\sum_{\textbf{k}',\alpha,\beta}\hat{c}^{\dagger}_{\alpha}(\textbf{k}'-\textbf{k})\left(\bm{\sigma}\right)_{\alpha\beta}\hat{c}_{\beta}(\textbf{k}')
\end{eqnarray}
which is the momentum space representation of the spin density operator of the conduction electrons. In the above equation, $\bm{\sigma}=(\sigma_{1},\sigma_{2},\sigma_{3})^{T}$, is the vector whose components are the Pauli matrices. The Latin indices, $\left\{i,j,k,...\right\}$, refer to space indices and the Greek indices, $\left\{\alpha,\beta,...\right\}$, refer to spin indices. Furthermore, the operators $\hat{c}_{\alpha}(\textbf{k})$, $\hat{c}^{\dagger}_{\alpha}(\textbf{k})$ are electron annihilation and creation operators, which satisfy the algebra,
\begin{eqnarray}
\{\hat{c}_{\alpha}(\textbf{k}_{1}),\hat{c}^{\dagger}_{\beta}(\textbf{k}_{2})\}=\delta_{\alpha\beta}\delta^ {3}(\textbf{k}_{1}-\textbf{k}_{2}),\\
\{\hat{c}_{\alpha}(\textbf{k}_{1}),\hat{c}_{\beta}(\textbf{k}_{2})\}=\{\hat{c}^{\dagger}_{\alpha}(\textbf{k}_{1}),\hat{c}^{\dagger}_{\beta}(\textbf{k}_{2})\}=0.
\end{eqnarray}
We will now consider the path integral representation of the reduced density matrix associated with the system as we did in section \ref{sec:I- spin-phonon theory}. As so, we are going to use a basis of coherent-states for the system and the reservoir (in this case we have to use Grassmanian variables for fermions), so that the states of the Hilbert space are written 
in the form,
\begin{eqnarray}
\ket{\textbf{S}}\otimes\hket{\gamma},
\end{eqnarray}
where,
\begin{eqnarray}
\ket{\textbf{S}}=\prod_{\textbf{k}}\ket{\textbf{S}(\textbf{k})}=\prod_{\textbf{k}}\hat{\mathcal{D}}(\textbf{S}(\textbf{k}))\ket{0},\\
\hket{\gamma}=\prod_{\alpha,\textbf{k}}\ket{\gamma_{\alpha}(\textbf{k})}=\prod_{\alpha,\textbf{k}}\exp\left(\hat{c}^{\dagger}_{\alpha}(\textbf{k})\gamma_{\alpha}(\textbf{k})\right))\ket{0}.
\end{eqnarray}
These states $\ket{\textbf{S}}$ are to be defined in such a way that they satisfy $\bra{\textbf{S}}\hat{\textbf{S}}(\textbf{k})\ket{\textbf{S}}=\textbf{S}(\textbf{k})\bra{\textbf{S}}\hat{\mathbbm{1}}_{S}\ket{\textbf{S}}$. Let us now consider the vector
\begin{eqnarray}
\ket{\textbf{S}'}=\prod_{i}(1+|\zeta(\textbf{S}_{i})|^{2})^{-j}\exp(\zeta(\textbf{S}_{i})\hat{S}_{-,i})\ket{\psi_{0}},
\end{eqnarray}
where $\ket{\psi_{0}}$ is the tensor product of highest weight vectors of the spin-$j$ representation and $\zeta(\textbf{S}_{i})=\tan(\theta_{i}/2)e^{i\varphi_{i}}$ denotes the stereographic projection of $\textbf{S}_{i}=j(\sin\theta_{i}\cos\varphi_{i},\sin\theta_{i}\sin\varphi_{i},\cos\theta_{i})^{T}$ through the North Pole. Clearly, this state satisfies
\begin{eqnarray}
\bra{\textbf{S}'}\hat{\textbf{S}}_{i}\ket{\textbf{S}'}=\textbf{S}_{i}\bra{\textbf{S}'}\hat{\mathbbm{1}}_{S}\ket{\textbf{S}'}=\textbf{S}_{i},
\end{eqnarray}
we can now replace, in the last equation, $\hat{\textbf{S}}_{i}$ and $\textbf{S}_{i}$ by their Fourier representations, yielding
\begin{eqnarray}
\bra{\textbf{S}'}\sum_{\textbf{k}}e^{i\textbf{k}\cdot\textbf{x}_{i}}\hat{\textbf{S}}(\textbf{k})\ket{\textbf{S}'}=\sum_{\textbf{k}}e^{i\textbf{k}\cdot\textbf{x}_{i}}\textbf{S}(\textbf{k}),
\end{eqnarray}
or,
\begin{eqnarray}
 \sum_{\textbf{k}}e^{i\textbf{k}\cdot\textbf{x}_{i}}\bra{\textbf{S}'}\left(\hat{\textbf{S}}(\textbf{k})-\textbf{S}(\textbf{k})\right)\ket{\textbf{S}'}=0,
\end{eqnarray}
Since this relation is valid for all $\textbf{x}_{i}$ we identify $\ket{\textbf{S}'}$ with $\ket{\textbf{S}}$ because, indeed, one has
\begin{equation}
\bra{\textbf{S}'}\hat{\textbf{S}}(\textbf{k})\ket{\textbf{S}'}=\textbf{S}(\textbf{k})\bra{\textbf{S}'}\hat{\mathbbm{1}}_{S}\ket{\textbf{S}'}.
\end{equation}
Assuming that at time $t_{0}$ equation \eqref{eq:Initial condition} holds, we easily find, in analogy with section \ref{sec:I- spin-phonon theory}, that the expression for the propagator is the same as \eqref{eq:Superpropagator} but now in $I[\textbf{S}_{1},\textbf{S}_{2}]$ one must sum over all spin-$j$ degrees of freedom and the Feynman-Vernon functional is now given by
\begin{widetext}
\begin{eqnarray}
\mathcal{F}[\textbf{S}_{1},\textbf{S}_{2}]=\int d\mu(\gamma_{0})d\mu(\gamma_{1})d\mu(\gamma_{2})\ k(-\gamma_{0}^{\dagger},t;\gamma_{1},t_{0}|\textbf{S}_{1})\times Z_{R}^{-1}k(\gamma_{1}^{\dagger},-i\beta;\gamma_{2},0|0)\times k^{*}(\gamma_{0}^{\dagger},t;\gamma_{2},t_{0}|\textbf{S}_{2}),
\label{eq:Feynman-Vernon electrons}
\end{eqnarray} 
\end{widetext}
where $d\mu(\gamma)=\prod_{\alpha,\textbf{k}}d^{2}\gamma_{\alpha}(\textbf{k})\exp\left(-\gamma_{\alpha}^{*}(\textbf{k})\gamma_{\alpha}(\textbf{k})\right)$ and the kernel $k(\gamma_{f}^{\dagger},t_{f};\gamma_{i},t_{i}|\textbf{S})$ is defined in the same way as in \eqref{eq:kernel k} (taking special care because Grassmannian variables anti-commute). The last expression assumes, as in \ref{sec:I- spin-phonon theory}, that the density matrix decouples at an initial time $t_{0}$ and that the reduced density matrix of the bath, at that time, takes the form of a Boltzmann factor with Hamiltonian $\hat{\mathcal{H}}_{R}$ and inverse temperature $\beta$. The minus sign in the first kernel in \eqref{eq:Feynman-Vernon electrons} is due to the anti-periodic conditions on the trace formula using fermionic coherent states. Unlike the case of the phonons we cannot compute the Gaussian integrals exactly. We would then like to do some expansion depending on the parameter $\lambda$. In order to achieve that we will perform a Hubbard-Stratonovich transformation and then expand a determinant resulting from the functional integrals. To do this, one can define an auxiliary bilinear form
\begin{widetext}
\begin{eqnarray}
\left(G^{-1}\right)_{\alpha\beta}(t-t',\textbf{k}-\textbf{k}';\textbf{S})
=-\frac{1}{2}\delta(t-t')\left[\bm{\sigma}_{\alpha\beta}\cdot\textbf{h}(\textbf{k}-\textbf{k}') 
+\lambda\bm{\sigma}_{\alpha\beta}\cdot\textbf{S}(t,\textbf{k}-\textbf{k}')\right].
\end{eqnarray}
With this definition, it is clear that we can do a Hubbard-Stratonovich transformation of the form
\begin{eqnarray}
\exp(-i\int_{t_{0}}^{t}\int_{t_{0}}^{t} ds ds'\sum_{\textbf{k}\textbf{k}',\alpha\beta}\gamma_{\alpha}^{*}(s,\textbf{k})\left(G^{-1}\right)_{\alpha\beta}(s-s',\textbf{k}-\textbf{k}';\textbf{S})\gamma_{\beta}(s',\textbf{k}'))=\\
=\int \frac{\prod_{\alpha,\textbf{k}} \mathcal{D}^{2}\zeta_{\alpha}(\textbf{k})}{\det \left(-iG\right) [\textbf{S}]}
\exp(i\int_{t_{0}}^{t}\int_{t_{0}}^{t} ds ds'\sum_{\textbf{k}\textbf{k'},\alpha\beta}\zeta_{\alpha}^{*}(s,\textbf{k})\left(G\right)_{\alpha\beta}(s-s',\textbf{k}-\textbf{k}';\textbf{S})\zeta_{\beta}(s',\textbf{k}')+ \\
+i\int_{t_{0}}^{t}\int_{t_{0}}^{t} ds\sum_{\textbf{k},\alpha}\zeta_{\alpha}^{*}(s,\textbf{k})\gamma_{\alpha}(s,\textbf{k})+\gamma_{\alpha}^{*}(s,\textbf{k})\zeta_{\alpha}(s,\textbf{k})).
\end{eqnarray}
\end{widetext}
Doing this, we achieve a linear coupling between the $\gamma$'s and the $\zeta$'s, which allows us to compute the Gaussian integrals in the $\gamma$'s. Replacing this transformation in the expression for the influence functional and computing the Gaussian integrals associated with the $\gamma$'s, we arrive at,
\begin{widetext}
\begin{eqnarray}
\int \frac{\prod_{\alpha,\textbf{k}}\mathcal{D}^{2}\zeta_{1,\alpha}(\textbf{k})}{\det(-iG)[\textbf{S}_{1}]}\frac{\prod_{\alpha,\textbf{k}}\mathcal{D}^{2}\zeta_{2,\alpha}(\textbf{k})}{\det(iG)[\textbf{S}_{2}]}\times\\
\times\exp\left(i\int_{t_{0}}^{t}\int_{t_{0}}^{t} ds ds'\sum_{\textbf{k}\textbf{k'},\alpha\beta} \left(\zeta_{1,\alpha}^{*}(s,\textbf{k}) \ \zeta_{2,\alpha}^{*}(s,\textbf{k})\right)
\left(\Delta(s-s',\textbf{k}-\textbf{k}')_{\alpha\beta}\right)\left(\begin{array}{c}
\zeta_{1,\beta}(s',\textbf{k}')\\
\zeta_{2,\beta}(s',\textbf{k}')
\end{array}\right)\right),
\end{eqnarray}
where,
\begin{eqnarray}
\left(\Delta(t,\textbf{k})_{\alpha\beta}\right)=\left(\begin{array}{cc}
\left(\mathcal{G}_{11}\right)_{\alpha\beta}(t,\textbf{k})+\left(G\right)_{\alpha\beta}(t,\textbf{k};\textbf{S}_{1}) & \left(\mathcal{G}_{12}\right)_{\alpha\beta}(t,\textbf{k})\\
\left(\mathcal{G}_{21}\right)_{\alpha\beta}(t,\textbf{k}) & \left(\mathcal{G}_{22}\right)_{\alpha\beta}(t,\textbf{k})-\left(G\right)_{\alpha\beta}(t,\textbf{k};\textbf{S}_{2})
\end{array}\right),
\end{eqnarray} 
in which,
\begin{eqnarray}
\left(\begin{array}{cc}
\left(\mathcal{G}_{11}\right)_{\alpha\beta}(t,\textbf{k}) & \left(\mathcal{G}_{12}\right)_{\alpha\beta}(t,\textbf{k})\\
\left(\mathcal{G}_{21}\right)_{\alpha\beta}(t,\textbf{k}) & \left(\mathcal{G}_{22}\right)_{\alpha\beta}(t,\textbf{k})
\end{array}\right)=\\
= 
i
\delta^{3}(\textbf{k})\delta_{\alpha\beta}e^{-i\epsilon(\textbf{k})t}\left(\begin{array}{cc}
\left(1-f(\epsilon(\textbf{k}))\right)\theta(t)-f(\epsilon(\textbf{k}))\theta(-t) & -f(\epsilon(\textbf{k}))\\
1-f(\epsilon(\textbf{k})) & \left(1-f(\epsilon(\textbf{k}))\right)\theta(-t) -f(\epsilon(\textbf{k}))\theta(t)
\end{array}\right),
\end{eqnarray}
with $f(x)=(e^{x}+1)^{-1}$ being the Fermi-Dirac distribution. 
\end{widetext}
The functional integral is readily evaluated to be,
\begin{eqnarray}
\det\{\left[-i\left(\begin{array}{cc}
\mathcal{G}_{11}+G(\textbf{S}_{1}) & \mathcal{G}_{12}\\
\mathcal{G}_{21} & \mathcal{G}_{22}-G(\textbf{S}_{2})
\end{array}\right)\right]\nonumber\\
\times\left[i\left(\begin{array}{cc}
G^{-1}(\textbf{S}_{1}) & 0\\
0 & -G^{-1}(\textbf{S}_{2})
\end{array}\right)\right]\}=\det(\mathbbm{1}+\mathfrak{S}),
\label{eq:Det Expansion}
\end{eqnarray}
where,
\begin{eqnarray}
\mathfrak{S}=\left(\begin{array}{cc}
\mathcal{G}_{11}G^{-1}(\textbf{S}_{1}) & -\mathcal{G}_{12}G^{-1}(\textbf{S}_{2})\\
\mathcal{G}_{21}G^{-1}(\textbf{S}_{1}) & -\mathcal{G}_{22}G^{-1}(\textbf{S}_{2})
\end{array}\right).
\end{eqnarray}
The determinant of equation \eqref{eq:Det Expansion} should be understood, of course, in the functional sense. We can write,
\begin{eqnarray}
\det(\mathbbm{1}+\mathfrak{S})&=&\exp\left\{\tr\left[\log(\mathbbm{1}+\mathfrak{S})\right]\right\}\nonumber\\
&=&\exp\left(\sum_{k}\frac{(-1)^{k+1}}{k}\tr{\mathfrak{S}^{k}}\right),
\end{eqnarray}
here the trace is also understood in the functional sense. In the way it is written, this produces an expansion in powers of the matrix elements of $\mathfrak{S}$ and consequently an expansion in the parameter $\lambda$.\\
\indent The first term of the expansion is easily found to be zero and if we keep only terms of second order in the matrix elements of $\mathfrak{S}$ we obtain, in the Keldysh representation of the fields as defined by equation \eqref{eq:Keldysh rep},
\begin{widetext}
\begin{eqnarray}
\log\left[\det\left(\mathbbm{1}+\mathfrak{S}\right)\right]&=&-\frac{1}{2}\tr\mathfrak{S}^{2}+\mathcal{O}(\mathfrak{S}^{3})\nonumber\\
&=&-\lambda\sum_{\textbf{k}}\int_{t_{0}}^{t}\int_{t_{0}}^{t} dt_{1}dt_{2}(\mathcal{S}g)(t_{1}-t_{2},\textbf{k})\theta(t_{1}-t_{2})\textbf{h}(-\textbf{k})\cdot\textbf{D}(t_{2},\textbf{k})\nonumber\\
&&-\lambda^{2}\sum_{\textbf{k}}\int_{t_{0}}^{t}\int_{t_{0}}^{t} dt_{1}dt_{2}(\mathcal{S}g)(t_{1}-t_{2},\textbf{k})\theta(t_{1}-t_{2})\textbf{S}(t_{1},-\textbf{k})\cdot\textbf{D}(t_{2},\textbf{k})\nonumber\\
&&-\frac{\lambda^{2}}{4}\sum_{\textbf{k}_{1}}\int_{t_{0}}^{t}\int_{t_{0}}^{t} dt_{1}dt_{2}(\mathcal{P}g)(t_{1}-t_{2},\textbf{k})\textbf{D}(t_{1},-\textbf{k})\cdot\textbf{D}(t_{2},\textbf{k})+\mathcal{O}(\mathfrak{S}^{3}),
\label{eq:Electron effective action}
\end{eqnarray}
\end{widetext}
where
\begin{eqnarray}
g(t,\textbf{k})&=&\sum_{\textbf{k}'}e^{-i\left(\epsilon(\textbf{k}')-\epsilon(\textbf{k}'-\textbf{k})\right)t}\nonumber \\
&&\times\left(1-f(\epsilon(\textbf{k}'))\right)f(\epsilon(\textbf{k}'-\textbf{k})),
\label{eq:def of g}
\end{eqnarray}
and
\begin{eqnarray}
\left(\mathcal{P}\phi\right)(\textbf{k})=\frac{1}{2}\left(\phi(\textbf{k})+\phi(-\textbf{k})\right),\nonumber \\ \left(\mathcal{S}\phi\right)(\textbf{k})=\frac{1}{2}\left(\phi(\textbf{k})-\phi(-\textbf{k})\right),
\end{eqnarray}
are the symmetrizer and the anti-symmetrizer operators associated with the momentum variable, respectively.
This makes it clear that the part leading to dissipation will be the term coupling two $\textbf{D}$ fields since it will introduce an imaginary term in the action. This is easy to see because the Fourier transform of an even (odd) function is real (imaginary), if the function is real.\\
\indent One can, analogously to what we did in section \ref{sec:I- spin-phonon theory}, introduce a random field using a Hubbard-Stratonovich transformation. This time the correlation function (in position space) reads
\begin{eqnarray}
\left\langle\xi_{\alpha}(t,\textbf{x})\xi_{\beta}(0)\right\rangle &=&\delta_{\alpha\beta}\frac{\lambda^{2}}{2}[(\mathcal{F}^{-1}\circ\mathcal{P})g](t,\textbf{x})\nonumber\\
&=&\delta_{\alpha\beta}\frac{\lambda^{2}}{2}\text{Re}\{[\mathcal{F}^{-1}g](t,\textbf{x})\},
\label{eq:spin-electron random-field correlation}
\end{eqnarray}
where $\mathcal{F}^{-1}$ denotes the inverse Fourier transform associated with the variable $\textbf{k}$. 
We now prove that a fluctuation-dissipation theorem holds for this effective theory. In order to do that, we integrate the correlation function of equation \eqref{eq:spin-electron random-field correlation},
\begin{widetext}
\begin{eqnarray}
\int_{t_{0}}^{t}ds\int d^{3}x\left\langle\xi_{\alpha}(s,\textbf{x})\xi_{\beta}(0)\right\rangle =\delta_{\alpha\beta}\frac{\lambda^{2}}{2}\text{Re}\{\int_{t_{0}}^{t}ds\int d^{3}x [\mathcal{F}^{-1}g](s,\textbf{x})\}.
\end{eqnarray}
Now we rewrite the integral appearing in the above expression by replacing the expression for $g(t,\textbf{k})$ of \eqref{eq:def of g},
\begin{eqnarray}
\int_{t_{0}}^{t}ds\int d^{3}x [\mathcal{F}^{-1}g](s,\textbf{x})&=& \int_{t_{0}}^{t}ds\int d^{3}x\sum_{\textbf{k}\textbf{k}'}e^{i\left[\textbf{k}\cdot\textbf{x}-i\left(\epsilon(\textbf{k}')-\epsilon(\textbf{k}'-\textbf{k})\right)s\right]}\nonumber \\
&&\times\left(1-f(\epsilon(\textbf{k}'))\right)f(\epsilon(\textbf{k}'-\textbf{k}))
\end{eqnarray}
\end{widetext}
Identifying the Dirac delta's appearing in the above expression, we find
\begin{eqnarray}
&&\sum_{\textbf{k}\textbf{k}'}\delta^{3}(\textbf{k})\delta\left(\epsilon(\textbf{k}')-\epsilon(\textbf{k}'-\textbf{k})\right)\nonumber \\
&&\times\left(1-f(\epsilon(\textbf{k}'))\right)f(\epsilon(\textbf{k}'-\textbf{k})), 
\end{eqnarray}
or, if we define the excitation energy $\omega(\textbf{k}',\textbf{k})=\epsilon(\textbf{k}'-\textbf{k})-\epsilon(\textbf{k}')$,
\begin{eqnarray}
&&\sum_{\textbf{k}\textbf{k}'}\delta^{3}(\textbf{k})\delta\left(\omega(\textbf{k}',\textbf{k})\right)\nonumber \\
&&\times\left(1-f(\epsilon(\textbf{k}'))\right)f(\epsilon(\textbf{k}')+\omega(\textbf{k}',\textbf{k})), 
\end{eqnarray}
The above formula has to be treated carefully. First we recall the useful identity,
\begin{equation}
f(\epsilon)(1-f(\epsilon -\omega))=n(\omega)(f(\epsilon-\omega)-f(\epsilon)),
\end{equation}
where $n(\omega)$ is the Bose-Einstein distribution, yielding
\begin{eqnarray}
&&\sum_{\textbf{k}\textbf{k}'}\delta^{3}(\textbf{k})\delta\left(\omega(\textbf{k}',\textbf{k})\right)\nonumber \\
&&\times n(\omega(\textbf{k}',\textbf{k}))\left(f(\epsilon(\textbf{k}'))-f(\epsilon(\textbf{k}')+\omega(\textbf{k}',\textbf{k}))\right), 
\end{eqnarray}
Because of the delta associated with the excitation energy we only need the integrand evaluated at $\omega(\textbf{k},\textbf{k}')=0$. Clearly, the expression is undetermined because the Bose-Einstein distribution diverges and the difference of Fermi-Dirac distributions goes to zero. To proceed we will expand the integrand for small $\omega(\textbf{k},\textbf{k}')$. Observing that $f'(\epsilon)=-f(\epsilon)(1-f(\epsilon))$ we can Taylor expand the part containing Fermi-Dirac distributions so that we obtain
\begin{eqnarray}
&&n(\omega(\textbf{k}',\textbf{k}))\left(f(\epsilon(\textbf{k}'))-f(\epsilon(\textbf{k}')+\omega(\textbf{k}',\textbf{k}))\right)\nonumber\\ &=&\frac{1}{\beta\omega(\textbf{k}',\textbf{k})}\left[\omega(\textbf{k}',\textbf{k})f(\epsilon(\textbf{k}'))\left(1-f(\epsilon(\textbf{k}'))\right)\right]\nonumber\\
&&+\mathcal{O}(\omega(\textbf{k}',\textbf{k}))\nonumber\\
&=&\frac{1}{\beta}f(\epsilon(\textbf{k}'))\left(1-f(\epsilon(\textbf{k}'))\right) +\mathcal{O}(\omega(\textbf{k}',\textbf{k})).
\end{eqnarray}  
This means that our sum becomes, simply,
\begin{eqnarray}
&&\frac{1}{\beta}\sum_{\textbf{k}\textbf{k}'}\delta^{3}(\textbf{k})\delta\left(\omega(\textbf{k}',\textbf{k})\right)\nonumber \\
&&\times f(\epsilon(\textbf{k}'))\left(1-f(\epsilon(\textbf{k}'))\right).
\end{eqnarray}
If we sum in $\textbf{k}'$ the only contribution of the sum will be that in which $\omega(\textbf{k}',\textbf{k})=0$, or $\epsilon(\textbf{k}'-\textbf{k})=\epsilon(\textbf{k})$. Assuming that $\epsilon(\textbf{k})$ is a power law in $|\textbf{k}|$ this implies that the only term surviving is the one with $\textbf{k}'=0$. This reasoning gives,
\begin{eqnarray}
&&\frac{1}{\beta}f(\epsilon(0))\left(1-f(\epsilon(0))\right)\sum_{\textbf{k}}\delta^{3}(\textbf{k})\nonumber\\&=&\frac{1}{\beta}f(\epsilon(0))\left(1-f(\epsilon(0))\right)=\frac{1}{4\beta}.
\end{eqnarray}
The final result is
\begin{eqnarray}
&&\int_{t_{0}}^{t}ds\int d^{3}x\left\langle\xi_{\alpha}(s,\textbf{x})\xi_{\beta}(0)\right\rangle \nonumber\\
&=&\int_{t_{0}}^{t}ds\int d^{3}x\text{Re}\left\{\left\langle\xi_{\alpha}(s,\textbf{x})\xi_{\beta}(0)\right\rangle\right\}\nonumber\\
&=& 2\alpha'_{G}k_{B}T\delta_{\alpha\beta},
\label{eq:Fdt e}
\end{eqnarray}
where we have defined $\alpha'_{G}=\lambda^{2}/16$. The above formula shows that a fluctuation-dissipation theorem holds. This should be expected because when we expanded the determinant and kept only second order terms we were, precisely, doing linear response theory.
%
\\
\\
\section{Composite model}
\label{sec:III- Collection of results}
We now collect the results from sections \ref{sec:I- spin-phonon theory} and \ref{sec:II- Spin-electron theory} in a theory modelling the interaction of a spin vector field with phonons and electrons. It is straightforward to see that the spin-phonon interaction of \eqref{eq:Spin-phonon interaction} generalizes to
\begin{eqnarray}
\hat{\mathcal{H}}_{i}&=&-\sum_{\textbf{k}}(H_{\alpha\mu}^{*}(\textbf{k})\hat{a}^{\dagger}_{\mu}(\textbf{k})\hat{S}_{\alpha}(\textbf{k}) \nonumber\\ &&+H_{\alpha\mu}(\textbf{k})\hat{S}_{\alpha}(-\textbf{k})\hat{a}_{\mu}(\textbf{k})),
\end{eqnarray}
in order to account for a spin vector field (each space position now has an independent spin-$j$ degree of freedom). The total Hamiltonian is now given by the Hamiltonian of section \ref{sec:II- Spin-electron theory} plus the phonon reservoir Hamiltonian and this interaction Hamiltonian.\\
Assuming again the factorization of the initial density matrix, the Feynman-Vernon functional of this model factorizes into the product of two functionals, one associated to the phonons and the other to the electrons,
\begin{eqnarray}
\mathcal{F}[\textbf{S}_{1},\textbf{S}_{2}]=\mathcal{F}_{p}[\textbf{S}_{1},\textbf{S}_{2}]\mathcal{F}_{e}[\textbf{S}_{1},\textbf{S}_{2}].
\end{eqnarray}
The functional associated with the electrons is approximated by the exponential of the effective action of \eqref{eq:Electron effective action} and the one associated with the phonons is given by
\begin{widetext}
\begin{eqnarray}
\mathcal{F}_{p}[\textbf{S},\textbf{D}]=\exp\left\{-i\int_{t_{0}}^{t}dt_{1}\int_{t_{0}}^{t}dt_{2}\sum_{\textbf{k}}\left(\sqrt{2}S_{\alpha}(t_{1},-\textbf{k})\ \frac{1}{\sqrt{2}}D_{\alpha}(t_{1},-\textbf{k})\right)\tilde{G}_{\alpha\gamma}(t_{1}-t_{2},\textbf{k})\left(\begin{array}{c}
\sqrt{2}S_{\gamma}(t_{2},\textbf{k})\\
\frac{1}{\sqrt{2}}D_{\gamma}(t_{2},\textbf{k})
\end{array}
\right)\right\},
\end{eqnarray}
where,
\begin{eqnarray}
i\tilde{G}_{\alpha\gamma}(t,\textbf{k})&=&\sum_{\mu}H_{\alpha\mu}^{*}(\textbf{k})e^{-i\omega_{\mu}(\textbf{k})t}\left(\begin{array}{cc}
0 & -\theta(-t)\\
\theta(t) & 1+2n(\omega_{\mu}(\textbf{k})) 
\end{array}
\right)H_{\gamma\mu}(\textbf{k})\nonumber\\
&=&\left(\begin{array}{cc}
0 & -(\Lambda)_{\alpha\gamma}(t,\textbf{k})\theta(-t)\\
(\Lambda)_{\alpha\gamma}(t,\textbf{k})\theta(t) & 2(\Lambda_{\beta})_{\alpha\gamma}(t,\textbf{k}) 
\end{array}
\right).
\end{eqnarray}
\end{widetext}
\indent Regarding the introduction of the random field in the equations of motion, now its correlations are given by the sum of two terms. The one arising from the electrons has been calculated in section \ref{sec:II- Spin-electron theory}. The contribution given by the phonons is given by
\begin{eqnarray}
&&\left\langle\xi_{\alpha}(t,\textbf{x})\xi_{\beta}(0)\right\rangle_{\text{phonons}}=\sum_{\textbf{k},\mu}e^{i\left(\textbf{k}\textbf{x}-\omega_{\mu}(\textbf{k})t\right)}\nonumber\\
&&\times H_{\alpha\mu}^{*}(\textbf{k})H_{\beta\mu}(\textbf{k})\coth\left(\frac{\beta\omega_{\mu}(\textbf{k})}{2}\right).
\end{eqnarray}
Assuming that $H_{\alpha\mu}(\textbf{k})=H_{\alpha}(\omega_{\mu}(\textbf{k}))$, this last expression can be written as
\begin{eqnarray}
&&\left\langle\xi_{\alpha}(t,\textbf{x})\xi_{\beta}(0)\right\rangle_{\text{phonons}}=\int_{0}^{\infty}d\omega \rho(\omega,\textbf{x})e^{-i\omega_{\mu}(\textbf{k})t}\nonumber\\
&&\times H_{\alpha}^{*}(\omega_{\mu}(\textbf{k}))H_{\beta}(\omega_{\mu}(\textbf{k}))\coth\left(\frac{\beta\omega}{2}\right),
\end{eqnarray}
in which
\begin{eqnarray}
\rho(\omega,\textbf{x})=\sum_{\textbf{k},\mu}\delta(\omega-\omega_{\mu}(\textbf{k}))e^{i\textbf{k}\cdot\textbf{x}}.
\end{eqnarray}
If we make the following additional assumption that
\begin{eqnarray}
\rho(\omega,\textbf{x})=\rho(\omega)\delta^{3}(\textbf{x}),
\end{eqnarray}
then the discussion of the simplified model of section \ref{sec:I- spin-phonon theory} applies to this part of the correlation function and we have, for the full correlation function,
\begin{eqnarray}
\text{Re}\left\{\left\langle\xi_{\alpha}(t,\textbf{x})\xi_{\beta}(0)\right\rangle\right\}&=&\delta_{\alpha\beta}(\alpha_{G}\varphi(t)\delta^{3}(\textbf{x})\nonumber\\&&+\frac{\lambda^{2}}{2}\text{Re}\left\{[\mathcal{F}^{-1}g](\textbf{x})\right\}),
\end{eqnarray}
in which we have applied the considerations we have made in section \ref{sec:I- spin-phonon theory} regarding the coupling constants. 
The discussion of the validity of the fluctuation-dissipation theorem is the same as in the end of section \ref{sec:I- spin-phonon theory} and \ref{sec:II- Spin-electron theory}. It is clearly satisfied by both contributions. 
\section{Conclusions}
\label{sec:Conclusions}
Even though the model introduced is very complex, we were able to obtain some interesting results. In the limit of high temperatures, the magnetic moments associated to the 3d electrons feel a random field which has two contributions: one from the interaction with the phonons which, with some additional assumptions regarding coupling constants, can be reduced to the one predicted by Brown \citep{Brown1963}; and the other which comes from the interaction with the conduction electrons. Besides an external magnetic field, the magnetic moments also feel an effective magnetic field associated with their interaction with the electrons. This effective magnetic field, given by \eqref{eq:extra magfield}, is related to the magnetic field felt by the electrons, $\textbf{h}(\textbf{k})$, and explicitly manifests the Fermi-Dirac statistics of these particles because it is weighted by the function $(\mathcal{S}g)(\textbf{k})=(1/2)[\sum_{\textbf{k}'}\left(1-f(\epsilon(\textbf{k}'))\right)f(\epsilon(\textbf{k}'-\textbf{k}))-(\textbf{k}\rightarrow -\textbf{k})]$, where $f$ is the Fermi-Dirac distribution. It is remarkable that one can obtain the Landau-Lifshitz-Bloch equation from the interaction with phonons using the results of Rebei \textit{et. al.}\cite{RebeiParker2003} and Garanin\cite{Garanin1997}. The limit of high temperatures should be further investigated in the case of the contribution given by the conduction electrons since it is not clear from the expression of the random field correlation function that it will have Brown's form, i.e. white noise, under some assumption of the conduction electrons' energy spectrum and density of states in such a limit. If this is case, then the Gilbert damping constant is given by the sum of two terms, one from the phonons and the other from the electrons. The one given by the conduction electrons appears to be independent of the electron density of states and is equal to $\lambda^{2}/16$.\\  
\indent The case of finite temperatures is described by two generalized functions which are essentially Fourier transforms of functions which characterize the type of interaction, the statistics and the density of states of the bath degrees of freedom. The part of the correlation function of the random field which comes from the interaction with the phonons yields an intimate relation between friction (the associated Gilbert constant) and the random field fluctuations, see \eqref{eq:Fdt ph}. Remarkably, this fluctuation-dissipation theorem is manifested even in the non-Markovian regime. The validity of the Markovian approximation is measured in terms of the thermal correlation time $\tau_{\text{Th}}$. For time scales lower than this, the Markovian approximation for the stochastic field fails. In the case of the contribution for the random field correlation function given by the electrons, we see that the equation \eqref{eq:Fdt e} manifests the existence of a fluctuation-dissipation theorem as it should due to the expansion done being the associated linear response theory. The theory considered here satisfies, thus, a general fluctuation-dissipation theorem which relates the random field fluctuations to the friction constants which measure the effect of the interaction of the spin with electrons and phonons. \\
\indent The model is more general than the so-called three temperature model considered and validated in experimental environment by Beaurepaire \textit{et. al.}\cite{Beaurepaire}. This is because we do not consider that the spins, electrons and phonons are thermalized. Instead, we consider that at an initial time $t_{0}$ they are thermalized and the density matrix decouples at that time, but the time evolution couples the various systems and, thus, considering individual temperatures at each instant has no precise meaning within this model.\\ 
\indent The model proposed here should be further investigated, namely numerically, since it might give a good description of the dynamics of magnetic moments in the case of ultrashort laser induced excitations of ferromagnetic materials at time scales below the picosecond.
\section*{Acknowledgements}
\label{Acknowledgements}
 We would like to acknowledge the financial support of the project PTDC/FIS/70843/2006, of Funda\c{c}\~{a}o para a Ci\^{e}ncia e a Tecnologia, Portugal, in particular the fellowship BL 486/2010, to BM.
%
\section*{Appendix: Equations of Motion}
\label{App:Equations of Motion}
The equations of motion in momentum space of the full theory resulting from taking the variation of the effective action (in which we consider the truncated expansion of \ref{sec:II- Spin-electron theory}) appearing in the path integral representation of the spin density matrix are
\begin{widetext}
\begin{eqnarray}
\dot{D}(s,\textbf{k})=\sum_{\textbf{p}}[\textbf{S}(s,\textbf{k}-\textbf{p})\times\textbf{W}(s,\textbf{p})+\textbf{D}(s,\textbf{k}-\textbf{p})\times\left(\textbf{T}(s,\textbf{p})+\textbf{H}_{\text{eff}}(s,\textbf{p})\right)],\\
\dot{S}(s,\textbf{k})=\sum_{\textbf{p}}[\textbf{S}(s,\textbf{k}-\textbf{p})\times\left(J\textbf{k}^{2}\textbf{S}(s,\textbf{p})+\textbf{T}(s,\textbf{p})+\textbf{H}_{\text{eff}}(s,\textbf{p})\right)+\frac{1}{4}\textbf{D}(s,\textbf{k}-\textbf{p})\times\left(
J\textbf{k}^{2}\textbf{D}(s,\textbf{p})+\textbf{W}(s,\textbf{p})\right)],
\end{eqnarray}
in which
\begin{eqnarray}
\textbf{T}(s,-\textbf{k})=\textbf{T}^{(S)}(s,-\textbf{k})+\textbf{T}^{(D)}(s,-\textbf{k}),\\
\textbf{T}^{(S)}(s,-\textbf{k})=\textbf{T}^{(S)}_{\text{ph}}(s,-\textbf{k})+\textbf{T}^{(S)}_{\text{sd}}(s,-\textbf{k}),\\
\textbf{T}^{(D)}(s,-\textbf{k})=\textbf{T}^{(D)}_{\text{ph}}(s,-\textbf{k})+\textbf{T}^{(D)}_{\text{sd}}(s,-\textbf{k}),\\
\textbf{T}^{(S)}_{\text{ph}}(s,-\textbf{k})=i\int_{t_{0}}^{t} ds'\theta(s'-s)\left[\Lambda(s-s',-\textbf{k})-\Lambda(s'-s,\textbf{k})\right]\textbf{S}(s',-\textbf{k}),\\
\textbf{T}^{(S)}_{\text{sd}}(s,-\textbf{k})=-i\lambda^{2}\int_{t_{0}}^{t} ds'(\mathcal{S}g)(-\textbf{k})\textbf{S}(s',-\textbf{k})\theta(s'-s),\\
\textbf{T}^{(D)}_{\text{ph}}(s,-\textbf{k})=i\int_{t_{0}}^{t} ds'\theta(s'-s)\left[\Lambda_{\beta}(s-s',-\textbf{k})-\Lambda_{\beta}(s'-s,\textbf{k})\right]\textbf{D}(s',-\textbf{k}),\\
\textbf{T}^{(D)}_{\text{sd}}(s,-\textbf{k})=-i\frac{\lambda^2}{2}\int_{t_{0}}^{t} ds'(\mathcal{P}g)(\textbf{k})\textbf{D}(s',-\textbf{k}),\\
\textbf{W}(s,-\textbf{k})=\textbf{W}_{\text{ph}}(s,-\textbf{k})+\textbf{W}_{\text{sd}}(s,-\textbf{k}),\\
\textbf{W}_{\text{ph}}(s,-\textbf{k})=i\int_{t_{0}}^{t} ds'\theta(s-s')\left[\Lambda(s-s',-\textbf{k})-\Lambda(s'-s,\textbf{k})\right]\textbf{D}(s',-\textbf{k}),\\
\textbf{W}_{\text{sd}}(s,-\textbf{k})=i\lambda^{2}\int_{t_{0}}^{t} ds'(\mathcal{S}g)(-\textbf{k})\textbf{D}(s',-\textbf{k})\theta(s-s'),\\
\textbf{H}_{\text{eff}}(s,-\textbf{k})=\textbf{H}(-\textbf{k})+i\lambda\int_{t_{0}}^{t} ds' (\mathcal{S}g)(\textbf{k})\textbf{h}(-\textbf{k}')\theta(s'-s),
\label{eq:extra magfield}
\end{eqnarray}
\end{widetext}
here the sub-indices ``ph'' and ``sd'' emphasize that this terms come from the interaction with phonon or with electrons, respectively. In the above expression the long wavelength approximation has been taken for the Heisenberg exchange term in the action so that it becomes a diffusive term in the equation of motion in which $J$ is some constant measuring the interaction strength.
\bibliographystyle{apsrev4-1.bst}
\bibliography{mybib}

\begin{thebibliography}{16}%
\makeatletter
\providecommand \@ifxundefined [1]{%
 \@ifx{#1\undefined}
}%
\providecommand \@ifnum [1]{%
 \ifnum #1\expandafter \@firstoftwo
 \else \expandafter \@secondoftwo
 \fi
}%
\providecommand \@ifx [1]{%
 \ifx #1\expandafter \@firstoftwo
 \else \expandafter \@secondoftwo
 \fi
}%
\providecommand \natexlab [1]{#1}%
\providecommand \enquote  [1]{``#1''}%
\providecommand \bibnamefont  [1]{#1}%
\providecommand \bibfnamefont [1]{#1}%
\providecommand \citenamefont [1]{#1}%
\providecommand \href@noop [0]{\@secondoftwo}%
\providecommand \href [0]{\begingroup \@sanitize@url \@href}%
\providecommand \@href[1]{\@@startlink{#1}\@@href}%
\providecommand \@@href[1]{\endgroup#1\@@endlink}%
\providecommand \@sanitize@url [0]{\catcode `\\12\catcode `\$12\catcode
  `\&12\catcode `\#12\catcode `\^12\catcode `\_12\catcode `\%12\relax}%
\providecommand \@@startlink[1]{}%
\providecommand \@@endlink[0]{}%
\providecommand \url  [0]{\begingroup\@sanitize@url \@url }%
\providecommand \@url [1]{\endgroup\@href {#1}{\urlprefix }}%
\providecommand \urlprefix  [0]{URL }%
\providecommand \Eprint [0]{\href }%
\providecommand \doibase [0]{http://dx.doi.org/}%
\providecommand \selectlanguage [0]{\@gobble}%
\providecommand \bibinfo  [0]{\@secondoftwo}%
\providecommand \bibfield  [0]{\@secondoftwo}%
\providecommand \translation [1]{[#1]}%
\providecommand \BibitemOpen [0]{}%
\providecommand \bibitemStop [0]{}%
\providecommand \bibitemNoStop [0]{.\EOS\space}%
\providecommand \EOS [0]{\spacefactor3000\relax}%
\providecommand \BibitemShut  [1]{\csname bibitem#1\endcsname}%
\let\auto@bib@innerbib\@empty
\bibitem [{\citenamefont {Rebei}\ and\ \citenamefont
  {Parker}(2003)}]{RebeiParker2003}%
  \BibitemOpen
  \bibfield  {author} {\bibinfo {author} {\bibfnamefont {A.}~\bibnamefont
  {Rebei}}\ and\ \bibinfo {author} {\bibfnamefont {G.~J.}\ \bibnamefont
  {Parker}},\ }\href@noop {} {\bibfield  {journal} {\bibinfo  {journal} {Phys.
  Rev. B}\ }\textbf {\bibinfo {volume} {67}},\ \bibinfo {pages} {104434}
  (\bibinfo {year} {2003})}\BibitemShut {NoStop}%
\bibitem [{\citenamefont {{A. Rebei, W. N. G. Hitchon and G. J.
  Parker}}(2005)}]{RebeiParker2005}%
  \BibitemOpen
  \bibfield  {author} {\bibinfo {author} {\bibnamefont {{A. Rebei, W. N. G.
  Hitchon and G. J. Parker}}},\ }\href@noop {} {\bibfield  {journal} {\bibinfo
  {journal} {Phys. Rev. B}\ }\textbf {\bibinfo {volume} {72}},\ \bibinfo
  {pages} {064408} (\bibinfo {year} {2005})}\BibitemShut {NoStop}%
\bibitem [{\citenamefont {Garanin}(1997)}]{Garanin1997}%
  \BibitemOpen
  \bibfield  {author} {\bibinfo {author} {\bibfnamefont {D.~A.}\ \bibnamefont
  {Garanin}},\ }\href@noop {} {\bibfield  {journal} {\bibinfo  {journal} {Phys.
  Rev. B}\ }\textbf {\bibinfo {volume} {55}},\ \bibinfo {pages} {3050}
  (\bibinfo {year} {1997})}\BibitemShut {NoStop}%
\bibitem [{\citenamefont {Keldysh}(1965)}]{Keldysh1964}%
  \BibitemOpen
  \bibfield  {author} {\bibinfo {author} {\bibfnamefont {L.~V.}\ \bibnamefont
  {Keldysh}},\ }\href@noop {} {\bibfield  {journal} {\bibinfo  {journal} {Sov.
  Phys. JETP}\ ,\ \bibinfo {pages} {1018}} (\bibinfo {year}
  {1965})}\BibitemShut {NoStop}%
\bibitem [{\citenamefont {Lindblad}(1976)}]{Lindblad1976}%
  \BibitemOpen
  \bibfield  {author} {\bibinfo {author} {\bibfnamefont {G.}~\bibnamefont
  {Lindblad}},\ }\href@noop {} {\bibfield  {journal} {\bibinfo  {journal}
  {Commun. Math. Phys.}\ }\textbf {\bibinfo {volume} {48}},\ \bibinfo {pages}
  {119} (\bibinfo {year} {1976})}\BibitemShut {NoStop}%
\bibitem [{\citenamefont {{O. Chubykalo-Fesenko, U. Nowak, R. W. Chantrell and
  D. Garanin}}(2008)}]{Chubykalo2008}%
  \BibitemOpen
  \bibfield  {author} {\bibinfo {author} {\bibnamefont {{O. Chubykalo-Fesenko,
  U. Nowak, R. W. Chantrell and D. Garanin}}},\ }\href@noop {} {\bibfield
  {journal} {\bibinfo  {journal} {Phys. Rev. B}\ }\textbf {\bibinfo {volume}
  {74}},\ \bibinfo {pages} {094436} (\bibinfo {year} {2008})}\BibitemShut
  {NoStop}%
\bibitem [{\citenamefont {{Beaurepaire, E., J.-C. Merle, A. Daunois, and J.-Y.
  Bigot}}(1996)}]{Beaurepaire}%
  \BibitemOpen
  \bibfield  {author} {\bibinfo {author} {\bibnamefont {{Beaurepaire, E., J.-C.
  Merle, A. Daunois, and J.-Y. Bigot}}},\ }\href@noop {} {\bibfield  {journal}
  {\bibinfo  {journal} {Phys. Rev. Lett.}\ }\textbf {\bibinfo {volume} {76}},\
  \bibinfo {pages} {4250} (\bibinfo {year} {1996})}\BibitemShut {NoStop}%
\bibitem [{\citenamefont {{A. Kirilyuk, A. V. Kimel and T.
  Rasing}}(2010)}]{Kirilyuk2010}%
  \BibitemOpen
  \bibfield  {author} {\bibinfo {author} {\bibnamefont {{A. Kirilyuk, A. V.
  Kimel and T. Rasing}}},\ }\href@noop {} {\bibfield  {journal} {\bibinfo
  {journal} {Rev. Mod. Phys.}\ }\textbf {\bibinfo {volume} {82}},\ \bibinfo
  {pages} {2731} (\bibinfo {year} {2010})}\BibitemShut {NoStop}%
\bibitem [{\citenamefont {{B. Hillebrands, K. Ounadjela,
  eds.}}(2002)}]{Hillebrands}%
  \BibitemOpen
  \bibfield  {author} {\bibinfo {author} {\bibnamefont {{B. Hillebrands, K.
  Ounadjela, eds.}}},\ }\href@noop {} {\emph {\bibinfo {title} {Spin Dynamics
  in Confined Magnetic Structures I}}},\ \bibinfo {edition} {1st}\ ed.\
  (\bibinfo  {publisher} {Springer-Verlag},\ \bibinfo {year}
  {2002})\BibitemShut {NoStop}%
\bibitem [{\citenamefont {{J. Xiao, A. Zangwill and M. D.
  Stiles}}(2006)}]{Xiao2005}%
  \BibitemOpen
  \bibfield  {author} {\bibinfo {author} {\bibnamefont {{J. Xiao, A. Zangwill
  and M. D. Stiles}}},\ }\href@noop {} {\bibfield  {journal} {\bibinfo
  {journal} {Phys. Rev. B}\ }\textbf {\bibinfo {volume} {72}},\ \bibinfo
  {pages} {014446} (\bibinfo {year} {2006})}\BibitemShut {NoStop}%
\bibitem [{\citenamefont {Vieira}\ and\ \citenamefont
  {Sacramento}(1995{\natexlab{a}})}]{Vieira1995-Ann}%
  \BibitemOpen
  \bibfield  {author} {\bibinfo {author} {\bibfnamefont {V.~R.}\ \bibnamefont
  {Vieira}}\ and\ \bibinfo {author} {\bibfnamefont {P.~D.}\ \bibnamefont
  {Sacramento}},\ }\href@noop {} {\bibfield  {journal} {\bibinfo  {journal}
  {Ann. Phys.}\ }\textbf {\bibinfo {volume} {242}},\ \bibinfo {pages} {188}
  (\bibinfo {year} {1995}{\natexlab{a}})}\BibitemShut {NoStop}%
\bibitem [{\citenamefont {Vieira}\ and\ \citenamefont
  {Sacramento}(1995{\natexlab{b}})}]{Vieira1995-Nucl}%
  \BibitemOpen
  \bibfield  {author} {\bibinfo {author} {\bibfnamefont {V.~R.}\ \bibnamefont
  {Vieira}}\ and\ \bibinfo {author} {\bibfnamefont {P.~D.}\ \bibnamefont
  {Sacramento}},\ }\href@noop {} {\bibfield  {journal} {\bibinfo  {journal}
  {Nucl. Phys. B}\ }\textbf {\bibinfo {volume} {448}},\ \bibinfo {pages} {331}
  (\bibinfo {year} {1995}{\natexlab{b}})}\BibitemShut {NoStop}%
\bibitem [{\citenamefont {Perelomov}(1985)}]{Perelomov}%
  \BibitemOpen
  \bibfield  {author} {\bibinfo {author} {\bibfnamefont {A.}~\bibnamefont
  {Perelomov}},\ }\href@noop {} {\emph {\bibinfo {title} {Generalized Coherent
  States and Their Applications}}}\ (\bibinfo  {publisher} {Springer-Verlag},\
  \bibinfo {year} {1985})\BibitemShut {NoStop}%
\bibitem [{\citenamefont {Gardiner}\ and\ \citenamefont
  {Zoller}(2000)}]{Gardiner2}%
  \BibitemOpen
  \bibfield  {author} {\bibinfo {author} {\bibfnamefont {C.~W.}\ \bibnamefont
  {Gardiner}}\ and\ \bibinfo {author} {\bibfnamefont {P.}~\bibnamefont
  {Zoller}},\ }\href@noop {} {\emph {\bibinfo {title} {Quantum Noise}}},\
  \bibinfo {edition} {2nd}\ ed.\ (\bibinfo  {publisher} {Springer-Verlag},\
  \bibinfo {year} {2000})\BibitemShut {NoStop}%
\bibitem [{\citenamefont {{G. W. Ford and R. F. O'Connel}}(1996)}]{Ford-1993}%
  \BibitemOpen
  \bibfield  {author} {\bibinfo {author} {\bibnamefont {{G. W. Ford and R. F.
  O'Connel}}},\ }\href@noop {} {\bibfield  {journal} {\bibinfo  {journal}
  {Nature}\ }\textbf {\bibinfo {volume} {113}} (\bibinfo {year}
  {1996})}\BibitemShut {NoStop}%
\bibitem [{\citenamefont {{W. F. Brown Jr.}}(1963)}]{Brown1963}%
  \BibitemOpen
  \bibfield  {author} {\bibinfo {author} {\bibnamefont {{W. F. Brown Jr.}}},\
  }\href@noop {} {\bibfield  {journal} {\bibinfo  {journal} {Phys. Rev.}\
  }\textbf {\bibinfo {volume} {130}},\ \bibinfo {pages} {1677} (\bibinfo {year}
  {1963})}\BibitemShut {NoStop}%
\end{thebibliography}%
\end{document}